\documentclass[english,twocolumn,prl]{revtex4}
\usepackage[T1]{fontenc}
\usepackage[latin9]{inputenc}
\usepackage{babel}
\usepackage{amsmath}
\usepackage{graphicx}
\usepackage{esint}
\usepackage{color}
\usepackage{xargs}[2008/03/08]

\makeatletter

\providecommand{\tabularnewline}{\\}

\@ifundefined{textcolor}{}
{%
 \definecolor{BLACK}{gray}{0}
 \definecolor{WHITE}{gray}{1}
 \definecolor{RED}{rgb}{1,0,0}
 \definecolor{GREEN}{rgb}{0,1,0}
 \definecolor{BLUE}{rgb}{0,0,1}
 \definecolor{CYAN}{cmyk}{1,0,0,0}
 \definecolor{MAGENTA}{cmyk}{0,1,0,0}
 \definecolor{YELLOW}{cmyk}{0,0,1,0}
}

\usepackage{braket}

\makeatother

\begin{document}

\title{Fate of many-body localization under periodic driving}

\author{Achilleas Lazarides$^{1}$, Arnab Das$^{2}$ and Roderich Moessner$^{1}$}
\affiliation{$^{1}$ Max-Planck-Institut f\"ur Physik komplexer Systeme, 01187 Dresden, Germany}
\affiliation{$^{2}$ Theoretical Physics Department, Indian Association for the Cultivation of Science, Kolkata 700032, India}

\newcommand{\tred}[1]{{\color{red} {#1}}}

\begin{abstract}

We study many-body localised quantum systems subject to periodic driving.
We find that the presence of a mobility edge \emph{anywhere} in the
spectrum is enough to lead to delocalisation for any driving strength
and frequency. By contrast, for a fully localised many-body system,
a delocalisation transition occurs at a finite driving frequency.
We present numerical studies on a system of interacting one-dimensional
bosons and the quantum random energy model, as well as simple physical pictures accounting for those results. 
\end{abstract}

\maketitle

\global\long\def\popdagger#1{f_{#1}^{\dagger}}
\global\long\def\pop#1{f_{#1}}

\begin{figure}
\includegraphics[scale=0.9]{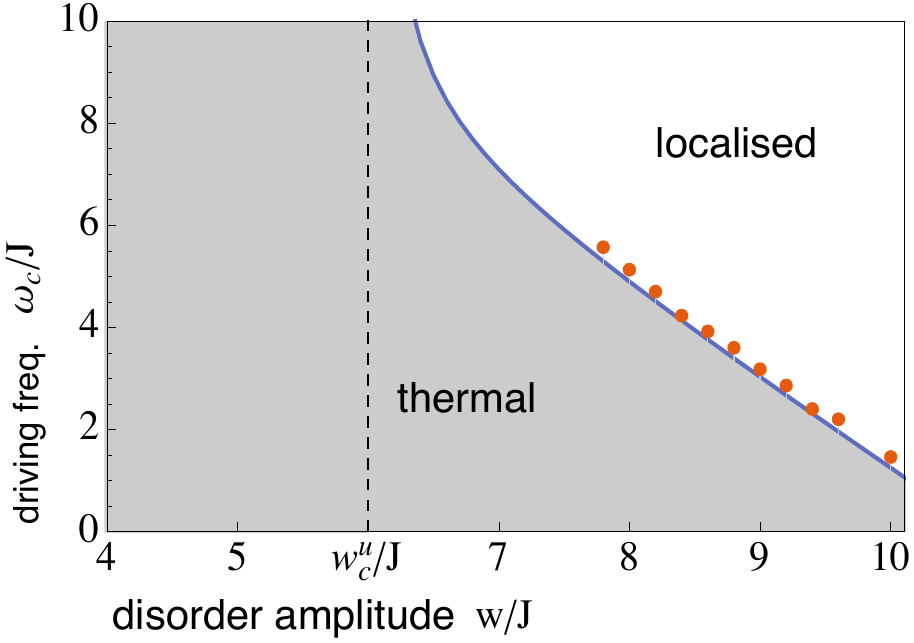}

\protect\caption{
Plot of driving frequency $\omega_{c}$ below which the system delocalises
as a function of disorder amplitude $w$. The shaded areas correspond
to delocalisation. The red dots are obtained from finite-size studies
of the level statistics of the system. The disorder amplitude $w_{c}$
is the value below which the undriven system is delocalised in the
absence of driving. The blue line is a guide to the eye. \label{fig:omegac-vs-w}}
\end{figure}

\begin{table}
	\begin{tabular}{|c|c|c|}
	\hline
	\textbf{Mobility edge} & low frequency & high frequency\tabularnewline
	\hline
	\hline
	present & delocalised & delocalised\tabularnewline
	\hline
	absent & delocalised & \textbf{localised}\tabularnewline
	\hline
	\end{tabular}

	\protect
	\caption{Effect of driving frequency in the presence
		and absence of a mobility edge \label{tab:Effect-of-driving}
	}

\end{table}

\emph{Introduction--}The study of disorder and localisation has a long and productive history, beginning with the seminal work of Anderson~\cite{Anderson1958}. More recently, the effects of disorder on interacting systems have been considered under the heading of many-body localisation (MBL)~\cite{Basko2006,Oganesyan2007}, in part motivated by fundamental questions relating to thermalisation in closed quantum systems.

At the same time, significant theoretical effort has been devoted to understanding thermalisation in periodically driven systems. There has been work recently on the long-time behaviour of both integrable~\cite{Lazarides:2014cl,Russomanno2012} and non-integrable~\cite{DAlessio:2014uv,Lazarides:2014ie,Ponte:2014vj} (with Ref.~\cite{Ponte:2014vj} also studying \emph{locally} driven MBL systems). For clean systems or MBL systems in their delocalised phase, it has been found that driving leads to a state equivalent to a fully-mixed state, satisfying a special case of the Eigenstate Thermalisation Hypothesis (ETH, see~ \cite{Deutsch:1991ju,PhysRevE.50.888,Rigol2008b,PhysRevE.89.042112,Lazarides:2014ie,DAlessio:2014uv,Ponte:2014vj}). Local periodic driving of MBL systems in their localised phase, on the other hand, has been argued not to have any global effects~\cite{Ponte:2014vj}.

In this work, we study the effects of \emph{global} periodic driving, and find that there exists a regime where MBL survives. We identify two mechanisms by which periodic driving might destroy MBL, depending on the existence or otherwise of a mobility edge. The first, rather robust, mechanism is the mixing of undriven eigenstates from everywhere in the spectrum by the driving; if there is a mobility edge, this results in delocalisation of all states of the effective Hamiltonian. The second mechanism is more subtle and involves strong mixing of states~\cite{Lazarides:2014ie} which cause a delocalisation transition at finite frequency.  Our findings are summarised in Table~\ref{tab:Effect-of-driving}.

In what follows, we begin by studying the case of no mobility edge. We introduce and numerically solve a system described by a local non-integrable Hamiltonian. After establishing the existence of the aforementioned critical frequency using level statistics, we demonstrate that ETH is (is not) satisfied below (above) this frequency, and present a physical picture explaining this phenomenon. We then move to the case where a mobility edge exists. As a case study, we use the Quantum Random Energy Model (QREM) which has recently been shown to display a mobility edge. A direct numerical solution confirms that driving delocalises the entire spectrum, consistent with an intuitive argument we sketch. Finally, we point out open questions.

We shall concentrate throughout on systems described by Hamiltonians of the form
\begin{equation}
H(t)=H_{0}+H_{D}(t),\label{eq:H0-hd}
\end{equation}
so that their time evolution is described by an effective Hamiltonian
$H_{eff}\left(\epsilon\right)$ for each instant $\epsilon$ during
the period $T$, defined by
\begin{equation}
\exp\left(-iH_{eff}\left(\epsilon\right)T\right)=\mathcal{T}\exp\left(-i\int_{\epsilon}^{\epsilon+T}dt\: H(t)\right).\label{eq:defn-heff}
\end{equation}
Without loss of generality we set $\epsilon=0$ (see Ref.~\cite{Lazarides:2014cl}). The eigenvalues of $H_{eff}\left(\epsilon\right)$, called the quasienergies, are independent of $\epsilon$ and effectively play the role of energy eigenvalues.

We now define what we mean by localised and delocalised phases. In a \emph{localised} phase, the (quasi-)energy level statistics do not display level repulsion, and the expectation values of operators in the eigenstates of the (effective) Hamiltonian fluctuate wildly from eigenstate to eigenstate. In a \emph{delocalised} phase, the opposite is true: the levels repel each other, and the expectation values of physical, local operators in nearby energy or quasienergy states are similar. Other definitions are possible and, in general, equivalent (see eg Ref.~\cite{PhysRevLett.113.107204}). The connection between the eigenstates of $H_{eff}$ and the applicability of ETH was elucidated in Ref.~\cite{Lazarides:2014ie}. The framework developed there turns out to be natural for discussing the case of a system which, in the absence of driving, is in the MBL phase.

\begin{figure}
	\includegraphics[scale=.95]{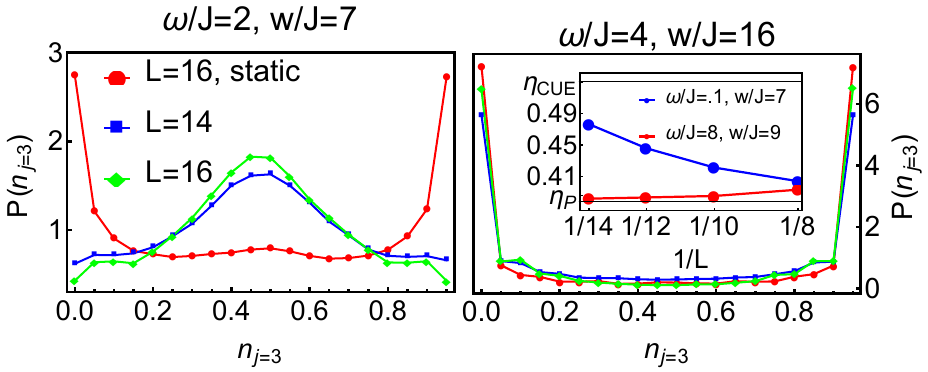}

	\protect\caption{
	Main plot: 
	Probability distribution of the eigenstate expectation values (EEVs) of the 
	density at site $j=3$. 
	Left (right): Driving with a low (high) frequency (see Fig.~\ref{fig:omegac-vs-w}) results in the probability distributions which does (does not) develop a central peak upon increasing system size, signalling delocalisation~\cite{Lazarides:2014ie}. Data is disorder averaged over $10^4$ (100) realisations for $L=14$ ($L=16$). Inset: Level statistics parameter versus inverse system size in the localised (bottom, red) delocalised (top, blue) phases. The parameter $\eta=\int ds\,s P(s)$ with $P(s)$ the probability distribution of the level statistics \cite{Oganesyan2007,DAlessio:2014uv}, taking the value $\eta_{P/CUE}$ in the localised/delocalised regime. Data averaged over $1000$ disorder realisations for $L=8, 10, 12$, $100$ realisations for $L=14$. 
	\label{fig:Plots-of-EEVs}
	}

\end{figure}

\emph{No mobility edge: Local Model--}Let us introduce a model of interacting hard-core bosons described by a driven, local Hamiltonian (Eq.~\ref{eq:H0-hd}) with
\begin{equation}
	H_{0} =
		H_{hop}
		+\sum_{r=1}^{2} 
			V_{r}\sum_{i=1}^{L-1}n_{i}n_{i+r}
		+\sum_{i=1}^{L}U_{i}n_{i}\label{eq:H}
\end{equation}
where $H_{hop}=\left(-\frac{1}{2}J\sum_{i=1}^{L-1}\left(b_{i}^{\dagger}b_{i+1} + b_{i+1}^{\dagger}b_{i} + hc \right)\right)$ is a hopping operator, the $b$ are hard-core bosonic operators, $U_{i}$ an on-site random potential uniformly distributed between $-w$ and $+w$ and $H_{D}\left(t\right)$ a time-periodic hopping term
\begin{equation}
	H_{D}\left(t\right)=\delta\tilde{\delta}(t) H_{hop}
	\label{eq:periodic-hopping}
\end{equation}
with $\delta$ a dimensionless constant, $\tilde{\delta}(t)=-1(+1)$ in the first (second) half of each period $T=2\pi/\omega$ . Via Jordan-Wigner transformations this model is related to a fermionic interacting system as well as to a spin-1/2 chain. Throughout this work we will concentrate on the specific case $V_{1}/J=V_{2}/J=1$, although our qualitative conclusions are not sensitive to this.

To locate the transition in the undriven model we use the standard technique~\cite{Oganesyan2007} involving finite-size scaling of the level statistics (see inset of Fig.~\ref{fig:Plots-of-EEVs} and Supplemental Material). At half-filling there thus appears to be a transition at a disorder amplitude $w_{c}^{u}/J$ ($\approx6$ for our interaction parameters $V_{1}/J=V_{2}/J=1$)~ \footnote{Energy-resolved level statistics (data not shown) does not indicate the existence of a mobile region in the spectrum for this particular model. }.


We now drive this system $\delta\neq0$. The level statistics of the quasienergies of $H_{eff}$ (Eq.~\ref{eq:defn-heff}), show level repulsion in the clean limit~\cite{DAlessio:2014uv} but are found to cross freely (indicating localisation) in the MBL regime if driven locally, as reported in Ref.~\cite{Ponte:2014vj}. Here, we show that globally periodically driving the system in the MBL regime delocalises the system if the driving frequency is below a (system size independent) critical value. We argue that this is a consequence of the structure of the effective Hamiltonian for an MBL system~\cite{1408.4297,1407.8480,2013PhRvL.111l7201S,Imbrie:2014vo,Ros:2014vz}.

As established above, the undriven system is in the delocalised phase for disorder amplitude $w<w_{c}^{u}$; driving at this disorder is qualitatively similar to driving any nonintegrable system~\cite{Lazarides:2014ie}, a case that has been studied in Ref.~\cite{Lazarides:2014ie}. We have indeed confirmed quasienergy level repulsion for $w<w_{c}^{u}$.

To study the MBL regime, $w>w_{c}^{u}$, we switch on periodic driving (Eq.~\ref{eq:periodic-hopping}) with amplitude $\delta/J=0.1$ (our results do not change qualitatively for different $\delta$ provided the system is large enough that the local level spacing is less than $\delta$). We directly calculate $H_{eff}$ and its level statistics. As our central result we find that for each disorder amplitude there exists a driving frequency $\omega_{c}\left(w\right)$ above which the system remains in the localised phase under driving (see Supplemental Material), while for $\omega<\omega_{c}\left(w\right)$ the system delocalises. This frequency is plotted in Fig.~\ref{fig:omegac-vs-w} as a function of disorder amplitude $w$, while examples of the level statistics results are shown in the inset of Fig.~\ref{fig:Plots-of-EEVs}. We expect $\omega_{c}\left(w\right)$ to diverge as $w$ approaches $w_{c}^{u}$ from above.

Having established a transition via the level statistics, we now show in addition that the phases above (below) $\omega_{c}$ do (do not) satisfy the form of ETH discussed in Ref.~\cite{Lazarides:2014ie}, further reinforcing our interpretation of $\omega_{c}$ as a ``delocalisation frequency''. We consider a localised undriven system and provide in  Fig.~\ref{fig:Plots-of-EEVs} direct evidence for the fully-mixed nature of the \emph{eigenstates} of $H_{eff}$ for slow -- but not for fast -- driving. The quantity under consideration is the probability distribution for the eigenstate expectation values (EEVs)~\cite{Lazarides:2014ie} of the density operator. Driving faster than the delocalisation frequency (right panel) yields little change in the probability distribution. By contrast, driving slowly (left panel), a central peak is seen to develop with increasing system size, corresponding to the EEVs all being equal and given by $n_{j=3}=0.5$. This is  the fully-mixed result for our system at half-filling, corresponding to delocalisation~\cite{Lazarides:2014ie}.

\begin{figure}
	\includegraphics[scale=0.65]{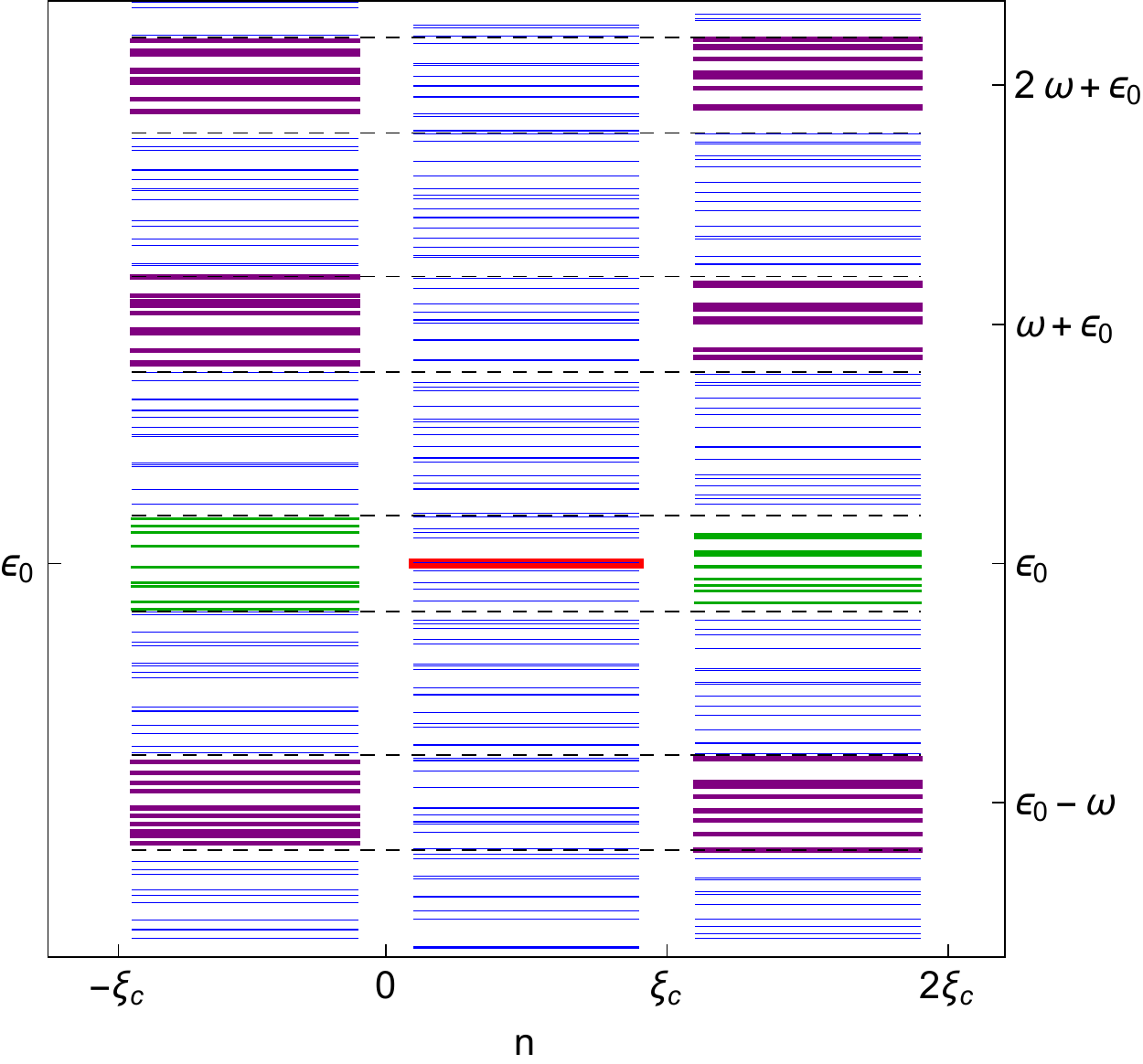}\\ 
	\protect\caption{
	Schematic illustration of sub systems and their energy levels in an MBL system. The horizontal axis indexes the conserved quantity (increasing $n$ corresponds to increasing spatial index $i$); the vertical indexes energy. $\xi_{c}$ is some localisation length, setting the typical spatial size of the subsystems. A periodic coupling of the subsystems with frequency $\omega$ couples the red state in the middle block to both the purple and the green levels in the two neighbouring blocks, while a time-independent coupling would only couple it to the green ones. The width of the purple and green strips is set by the amplitude of the driving. Critically, the limit of $\omega$ greater than the typical subsystem bandwidth is indistinguishable from a time-independent driving. By contrast, the limit of $\omega\rightarrow0$ collapses the local spectra, wiping out the effect of disorder.\label{fig:blocks}} 
\end{figure}

In passing, let us remark that, since the EEVs of the instantaneous Hamiltonian show the same behaviour as in Fig.~\ref{fig:Plots-of-EEVs} (data not shown), our localised phase is not unlike the localisation in energy space discussed in Ref.~\cite{Dalessio:2013}, even though the underlying physics is quite different.

\emph{Physical picture--}We now relate our numerical findings to a physical picture valid for weak driving. In the MBL phase and in the absence of driving, the system is effectively integrable in that there exist extensively many local integrals of motion~\cite{2013PhRvL.111l7201S,Ros:2014vz,1407.8480,1408.4297,Imbrie:2014vo}. The system may thus be thought of as a set of \emph{local} subsystems, of finite spatial extend, therefore of finite energetic bandwidth as schematically shown in Fig.~\ref{fig:blocks}. As a result, if the driving frequency is larger than the typical local subsystem bandwidth, the system cannot absorb energy from the driving and does not react. Therefore, \emph{driving with a frequency much higher than the typical local bandwidth cannot destroy MBL}. In contrast low-frequency driving may be understood by viewing our driving protocol as a series of quenches: as MBL systems eventually reach a steady-state after an instantaneous quench~\cite{SerbynQuench,2014arXiv1407.4476V}, periodic driving with the protocol we use can be thought of as a series of non-adiabatic perturbations. It is quite natural then to expect this to cause the system to spread in energy space, delocalising it.

Let us elaborate this pair of arguments, beginning with high-frequency driving.

\emph{High-frequency driving--}The most general form of $H_{MBL}$ consistent with known phenomenology such as vanishing of the conductivity at all energies is 
\begin{equation}
H_{MBL}^{G}=\sum_{n}\mathcal{{H}}_{n}^{\left(\ell\right)}+\sum_{m<n}\mathcal{{H}}_{n}^{\left(\ell\right)}V_{m,n}^{\left(\ell\right)}\mathcal{{H}}_{m}^{\left(\ell\right)}\ldots\label{eq:h-mbl-general}
\end{equation}
with the $\mathcal{{H}}_{n}^{\left(\ell\right)}$ Hamiltonians for local subsystems (with local spatial support) and $n$ a spatial index indicating the site about which the subsystem is centred~\cite{1408.4297,2013PhRvL.111l7201S,1407.8480,Imbrie:2014vo,Ros:2014vz}. Due to its locality, each $\mathcal{{H}}_{n}^{\left(\ell\right)}$ has a local spectrum of some typical, \emph{finite} width set by the disorder amplitude and other system details and independent of the other blocks (see Fig.~\ref{fig:blocks}, where the spectra for three $\mathcal{{H}}_{n}^{\left(\ell\right)}$ are sketched schematically). 

Driving $H_{MBL}^{G}$ with a sum of local terms such as in Eq.~\ref{eq:H0-hd} couples each $\mathcal{{H}}_{n}^{\left(\ell\right)}$ to its neighbours~ \footnote{Actually, due to the spatially exponentially localised nature of the conserved quantities, driving couples blocks at all instances by exponentially weakly; this however does not affect our argument.} via terms allowing energy and matter transfer. Consider a single energy level for $n=0$ (middle block, Fig.~\ref{fig:blocks}), indicated by the red line in the middle block. A time-independent coupling between the blocks couples it to the green blocks on each side, while a periodic coupling with frequency $\omega$ couples it to both the green and purple blocks by virtue of folding the energy spectrum into the $\omega$-periodic quasienergies. Crucially, for $\omega$ larger than the typical width of the blocks, folding the local spectra has no effect~\cite{Eckardt2005a} and a weak coupling \emph{does not} delocalise the system, as it acts similarly to a time-independent perturbation~\footnote{Note that we are only interested in perturbations much weaker than the disorder\textendash the opposite limit might be qualitatively different (see for example~\cite{1405.3966}).}. In other words, the system can react to the driving by absorbing energy quanta $\omega$ only if there exist levels separated by this energy. In the presence of MBL the typical local bandwidth sets the maximum driving frequency to which the system can react.\footnote{A
simple example of driving faster than this is given in the Supplemental
Material, which includes Refs.~\cite{Canovi2011,Grifoni1998,Das2010}.}


\emph{Low frequency--} In the limit of low-frequency driving disorder is effectively suppressed and the delocalised phase is always reached.

This phenomenon is best understood in the time domain as follows. Consider time evolving with Hamiltonian $H_{1(2)}$ for the first (second) half of the period. This series of nonadiabatic changes to the system generically results in a broadening of the energy distribution, provided that the half-period $T/2$ is longer than the characteristic relaxation time~\cite{2014arXiv1407.4476V,SerbynQuench}. Typically, this eventually leads to a fully-mixed state occupying the entire Hilbert space equiprobably.

There are two central ingredients to this argument. The first is that the relaxation time does not diverge with system size so that the half-period $T/2$ \emph{can} be longer. The existence of a dephasing timescale independent of system size~\cite{2014arXiv1407.4476V,SerbynQuench} ensures that this is the case. The second is that repeatedly dephasing in the two different eigenstate bases does lead to energy delocalisation. Since $H_{1,2}$ are both MBL Hamiltonians, the eigenstates of one are in general localised in terms of the eigenstates of the other. Nevertheless, repeated cycles of dephasing to alternating bases do indeed eventually lead to a fully-mixed state, as is shown in the Supplemental Material.

\begin{figure}
	\includegraphics[scale=0.9]{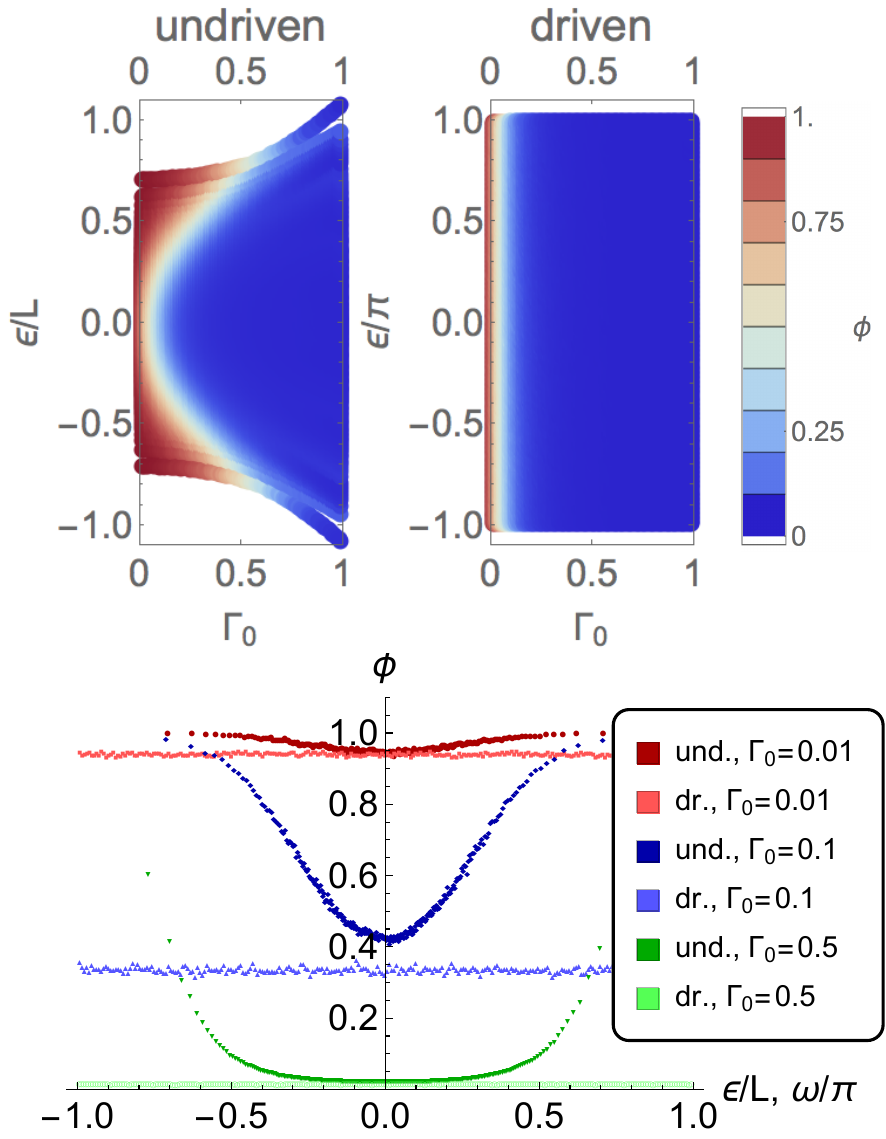}

	\protect\caption{Driving the QREM. The top left figure shows the participation ratio
	$\phi$ for the eigenstates of the undriven model, showing a mobile
	region (blue) surrounded by a localised region (red). Driving with
	frequency $\omega/J=0.1$ and amplitude $\delta/J=0.2$ (top right)
	causes all states at a given $\Gamma_{0}$ to become as delocalised
	as the least localised state at that $\Gamma_{0}$ in the undriven
	model. This is also shown in the bottom panel which shows $\phi$
	for $\Gamma_{0}=0.01,0.1,0.5$ (red, blue and green line, from top
	to bottom) in the absence (presence) of driving with darker (lighter)
	colour. The driven points always lie below the undriven points for
	the corresponding $\Gamma_{0}$. This is due to the strong mixing
	of all undriven eigenstates by the driving. All data in this figure
	is for $8$ spins and averaged over 1000 disorder realisations.\label{fig:QREM}}
\end{figure}

\newcommand{\sz}[1]{\sigma_{#1}^{z}}
\newcommand{\sx}[1]{\sigma_{#1}^{x}}

\emph{A mobility edge: The QREM as a case study--}We now turn to the case in which a mobility edge is present in the undriven spectrum. Our central result is based on the observation~\cite{Lazarides:2014ie} that a periodic perturbation acting on a system couples each undriven state to states spread uniformly throughout the spectrum of $H_{0}$. As a result, if part of the spectrum corresponds to delocalised eigenstates then all eigenstates of $H_{eff}$ will necessarily be delocalised. We numerically confirm this by studying the Quantum Random Energy Model (QREM), recently studied in Ref.~\cite{Laumann:2014vd} where it was shown to have a mobility edge. This model is described in Ref.~\cite{Laumann:2014vd}: it is defined for $N$ Ising spins with the Hamiltonian $H = E\left(\left\{\sz{j} \right\}\right) - \Gamma \sum_j \sx{j}$, where $E$ is a random operator diagonal in the $\sigma^z$ basis (that is, it assigns a random energy to each spin configuration) and $\Gamma$ a transverse field. Extensivity of the many-body spectrum is satisfied if the random energies are drawn from a distribution $P\left(E\right)=\frac{1}{\sqrt{\pi N}}\exp\left(-E^{2}/N\right)$.

The diagnostic of localisation we use is the participation ratio (PR), defined for the state $\ket\psi$ as $\phi=\sum_{n}\mbox{\ensuremath{\left|\left\langle n\right|\left.\psi\right\rangle \right|}}^{4}$ with $n$ enumerating Fock states. $\phi$ approaches unity for a state localised on a single Fock state and $2^{-N}$ for one fully delocalised in Fock space. The leftmost panel shows in Fig.~\ref{fig:QREM} shows the average $\phi$ versus energy (scaled with system size) of the 256 eigenstates of an undriven $N=8$ system averaged over 1000 disorder realisations, demonstrating the existence of a mobility edge.

Next, we drive the system by modulating $\Gamma\left(t\right)=\Gamma_{0}\left(1+\delta\tilde{\delta}\left(t\right)\right)$, $\tilde{\delta}\left(t\right)=+1(-1)$ for the first (second) half of the period with an amplitude $\delta=0.2$ and frequency $\omega=2\pi/T=0.1$. The PR of the eigenstates of $H_{eff}$ are shown in the second panel of Fig.~\ref{fig:QREM}. As expected, periodic driving causes delocalisation of the entire spectrum so long as part of the undriven spectrum at the same $\Gamma_{0}$ is delocalised.

\emph{Outlook--}We have shown that many-body systems can remain many-body localised, with Poissonian level statistics, when they are subjected to slow driving. On the other hand, for fast driving or in the presence of a mobility edge, delocalisation will occur, with driving inducing level repulsion.

This ``classification'' of the behaviour of MBL systems under driving immediately raises further questions. What are the timescales involved in reaching the long-time state we have discussed, how do they depend on the driving amplitude and frequency, and how do they differ between the localised and the delocalised limit? What is the precise difference between local and global driving as far as both the long-time state and the approach to it are concerned? More broadly, we have concentrated on systems with a bounded local spectrum. What happens if it is unbounded, as in the cases of a continuum system or of a lattice boson system? What if we bring the system in contact with a heat bath?

We believe that the dual out-of-equilibrium situation -- driving and MBL -- is only beginning to be explored and will prove to be fertile ground for future research.

\emph{Acknowledgements--}We thank J.~Bardarson, M.~Haque, V.~Khemani, J.~Kjall, V.~Oganesyan, S.~Sondhi, O.~Tieleman and particularly T.~Scheler for discussions as well as D.~Abanin, A.~Chandran and L.~D'Alessio for exchanges in the course of this project.

\emph{Note added:} After the completion of this work, two related works~\cite{Ponte2014b, AbaninDeRoeckHuveneers} have appeared. Each of these takes a somewhat different perspective but they all establish phenomenologies essentially consistent with the one we report.

\title{Supplemental material for 
	``The fate of many-body localization under periodic driving''
}

\section{Determination of the transition\label{sec:Determination-of-omegac}}

To accurately locate the localisation-delocalisation transition for
the undriven system, we first study the level statistics of the eigenvalues
of $H_{0}$, as for example in Ref.~\cite{Oganesyan2007}. That is,
after obtaining the (quasi-)energies $\epsilon_{n}$, we calculate
the following ratio involving adjacent level spacings $\delta_{n}=\epsilon_{n}-\epsilon_{n+1}$: $r_{n}=\min\left(\delta_{n},\delta_{n-1}\right)/\max\left(\delta_{n},\delta_{n-1}\right)$. The mean $\eta=\int_{0}^{1}dr\, rP(r)$
distinguishes between Wigner-Dyson-type and Poisson statistics. We
calculate $\eta$ for a sequence of system sizes and extrapolate the
limit of $\eta$ as $L\rightarrow\infty$.

\begin{figure*}
\includegraphics[scale=0.55]{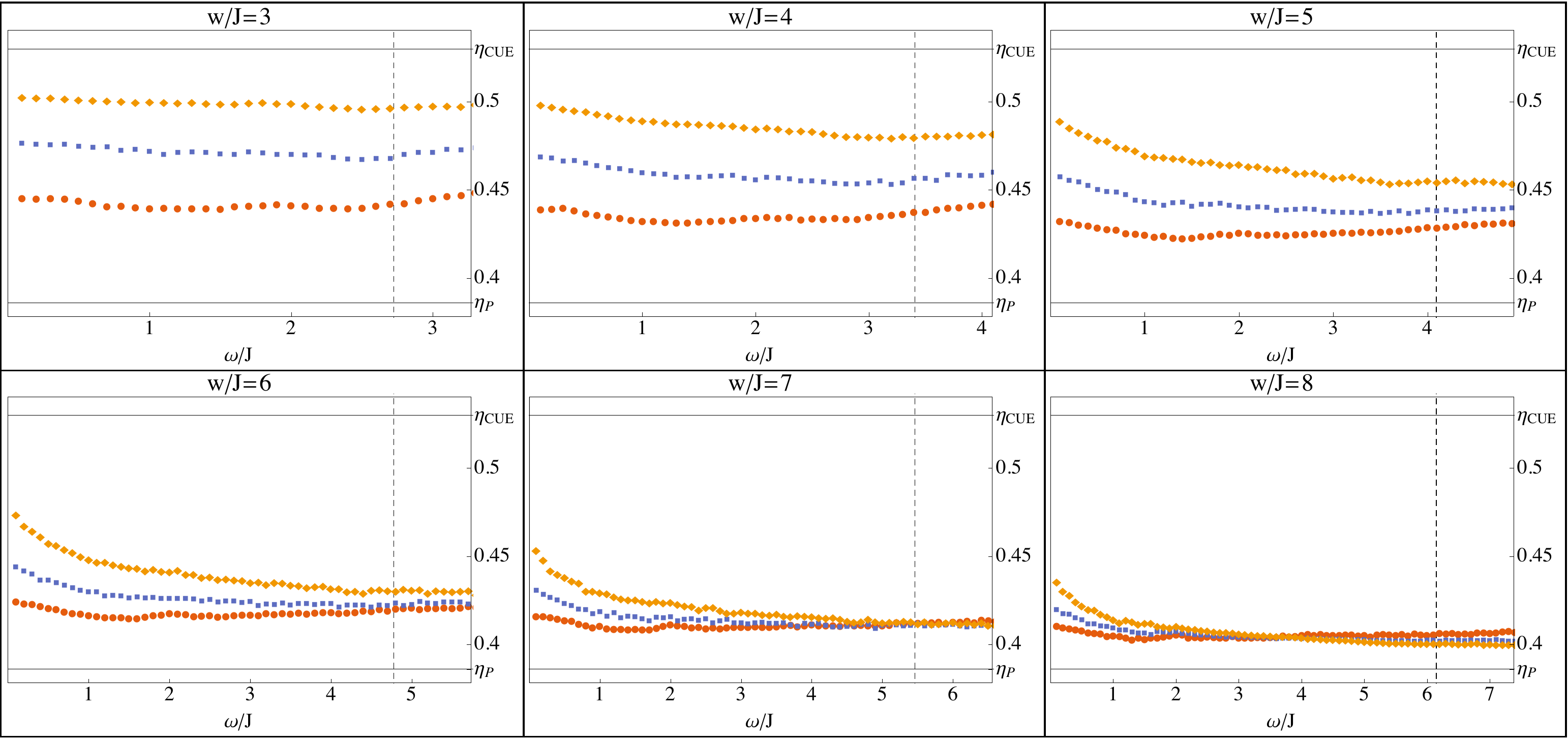}\protect\caption{Level statistics for various disorder amplitudes $w/J$ as a function
of driving frequency $\omega$. The driving amplitude is $\delta/J=0.1\ll w/J,\omega/J$,
and each point represents an average over 10000 disorder realisations.
The dashed vertical lines indicate half the width of the energy spectrum;
for $\omega$ greater than this, our results cannot be extrapolated
to the thermodynamic limit (see \ref{sec:Width-of-spectrum}). The
colours correspond to system sizes $L=8,10,12$ from bottom to top
for the smallest $\omega$. The values $\eta_{GOE}$ and $\eta_{P}$
correspond to the presence and absence of level repulsion, respectively,
which in turn correspond to localised and delocalised phases. The
dotted vertical lines correspond to the typical spectral width of
the system, for frequencies above which our results cannot be used
to infer the thermodynamic limit.\label{fig:example-level-stats}}
\end{figure*}

To obtain the frequency $\omega$ above which delocalisation sets in for a driven
system, we again calculate the mean of the distribution function of
the quasienergy statistics $\eta=\int_{0}^{1}dr\, P\left(r\right)$
as a function of disorder amplitude $w$, averaged over 10000 disorder
realisations and for several system sizes. Typical results are shown
in Fig.~\ref{fig:example-level-stats}. The transition is located
at the crossing point of the lines for different system sizes: if
increasing system size results in larger $\eta$ then we conclude
that the system is delocalised, since $\eta=\eta_{CUE}$ for a delocalised
system and $\eta=\eta_{P}$ for a localised system with $\eta_{P}<\eta_{CUE}$. 
Here $\eta_{CUE}$ is the value for the CUE ensemble~\cite{DAlessio:2014uv}.

To ensure that our results are applicable to the thermodynamic limit
we need to take a frequency low enough so that the width of the energy
spectrum of the undriven Hamiltonian is larger than the driving frequency
$\omega$. The main practical problem is the following: with decreasing
disorder amplitude $w$ and for fixed system size, the value $\omega_{c}$
increases while the energetic width of the DOS decreases (see Sec.~\ref{sec:Width-of-spectrum}).
Since $\omega$ must be small compared to the width in order for our
extrapolation to the thermodynamic limit to be meaningful, the $\omega_{c}$
for values of the disorder close to $w_{c}$ are inaccessible for
the system sizes available to us. The width of the DOS is indicated
in Fig.~\ref{fig:example-level-stats} by vertical lines; the crossing
point of the curves cannot lie to the right of this line, since otherwise
the finite size of the system would be important (and thus the results
would not be reliable in the thermodynamic limit).

Fig.~\ref{fig:example-level-stats} reveals the following
features: for $w/J\leq6$ (where the undriven system is delocalised)
the lines for succesive, increasing $L$ do not cross for values of
$\omega$ below the bandwidth, indicating that the thermodynamic limit
is delocalised, as expected. For $w/J>6$, there is a clear crossing
point, which indicates the position of the transition. The crossing
value of $\omega$ determined by this method is plotted as a function
of $w/J$ in Fig.~1 in the main text.

\section{Eigenstate expectation values} 
\label{sec:eigenstate_expectation_values}

As discussed in Ref.~\cite{Lazarides:2014ie} and the main text 
(Fig. 2 of the main text), 
periodically-driven ergodic (or delocalised) systems develop a peak
in the probability distribution of the eigenstate expectation values
(EEVs). Fig.~\ref{fig:Plots-of-EEVs} shows explicit examples of the EEVs in
the case of slow driving (system remains localised, left panel) and fast
driving (system delocalises, right panel).

\begin{figure}
\includegraphics[scale=0.4]{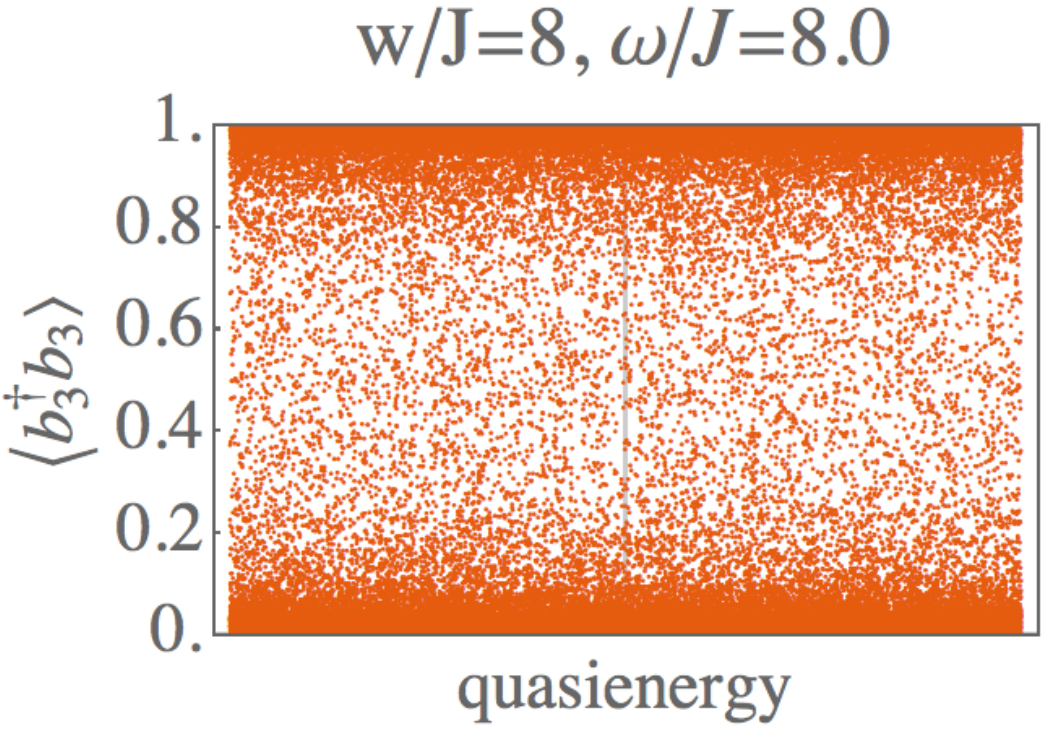}
\includegraphics[scale=0.4]{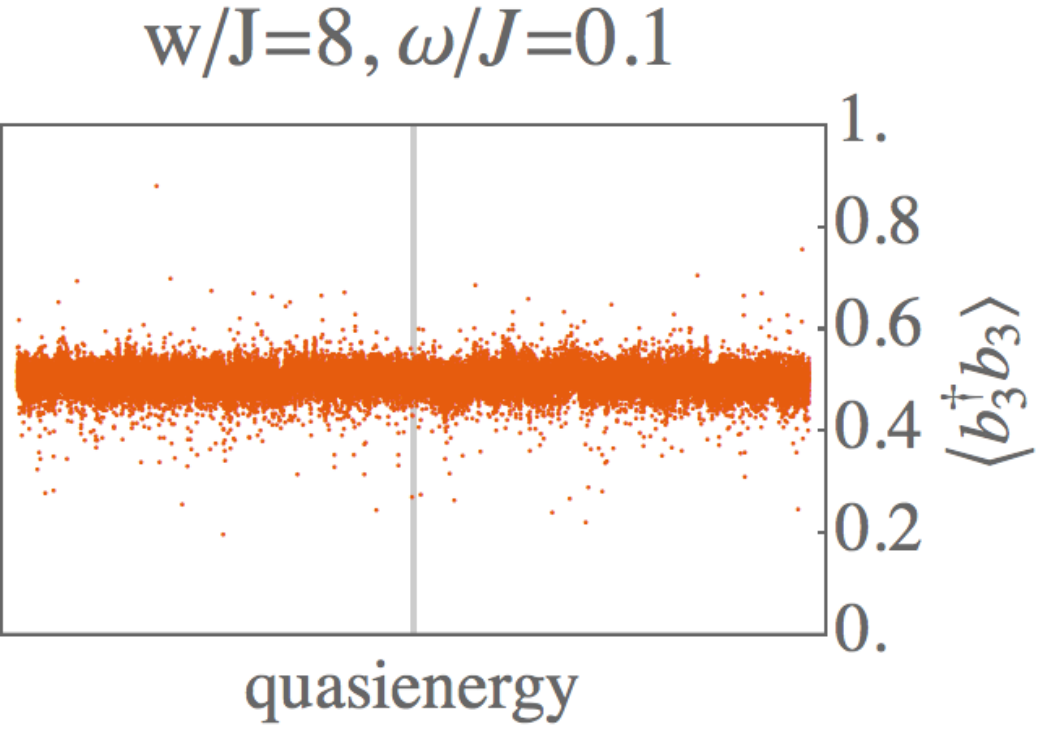}

\protect\caption{Plots of eigenstate expectation values (EEV) of the density at a single
arbitrarily chosen site in all the eigenstates of $H_{eff}$ for a
system with $w/J=8.0$, size $L=18$ for a Hilbert space dimension
of $D_{H}=48620$ and driving amplitude $\delta/J=0.1$. For driving
frequency above the blue line in Fig.~1 of the main text, $\omega/J=8.0$
(left), the EEVs fluctuate wildly between different eigenstates of
$H_{eff}$. In contrast, for a driving frequency below the blue line,
$\omega/J=0.1$ (right), there is markedly less eigenstate-to-eigenstate
variation, consistent with all states being fully mixed. This is the
expected behaviour of the EEVs for clean (therefore delocalised) driven
systems (see Ref.~\cite{Lazarides:2014ie}). In the undriven system
the EEVs appear qualitatively similar to those in the left panel.
\label{fig:Plots-of-EEVs}}
\end{figure}


\section{Single-particle localisation length} 
\label{sec:single_particle_localisation_length}
\begin{figure}
\includegraphics[scale=.5]{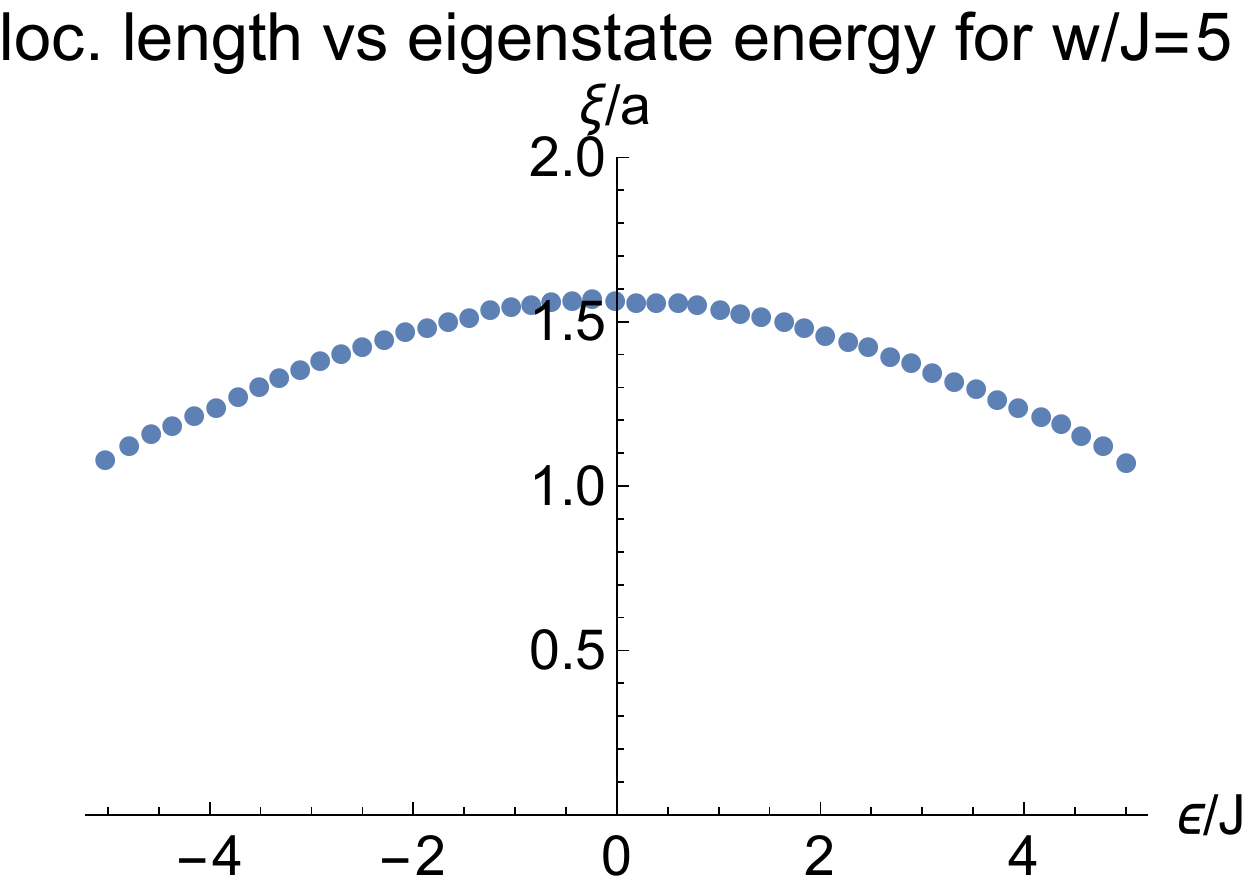}

\protect\caption{
	Single-particle localisation length for a noninteracting
	Anderson problem with parameters similar to the
	typical values used in our main text. The localisation length is well
	below most of the system sizes we have used.
	\label{fig:sp-loc-length}}
\end{figure}

In Fig.~\ref{fig:sp-loc-length} we show the single-particle localisation length
as a function of the eigenstate energy for the noninteracting Hamiltonian 
\begin{equation}
	H_{0} = 		
	-\frac{1}{2}J\sum_{i=1}^{L-1}
			\left(
			b_{i}^{\dagger}b_{i+1} + b_{i+1}^{\dagger}b_{i} +
			hc\right)
			+\sum_{i=1}^{L}U_{i}n_{i}
			\label{eq:H}
\end{equation}
with $U_{i}$ an on-site random potential uniformly distributed between $-w$ and
$+w$ and we take $w/J=5$. This plot demonstrates that, for the values of $w/J$ where the
interacting Hamiltonian of the main text is in the MBL phase, the single-particle
localisation length is well below the accessible system sizes. This ensures that 
the finiteness of this length is not a source of finite-size effects.

\section{Width of spectrum of local model\label{sec:Width-of-spectrum}}

\begin{figure}
\includegraphics[scale=0.75]{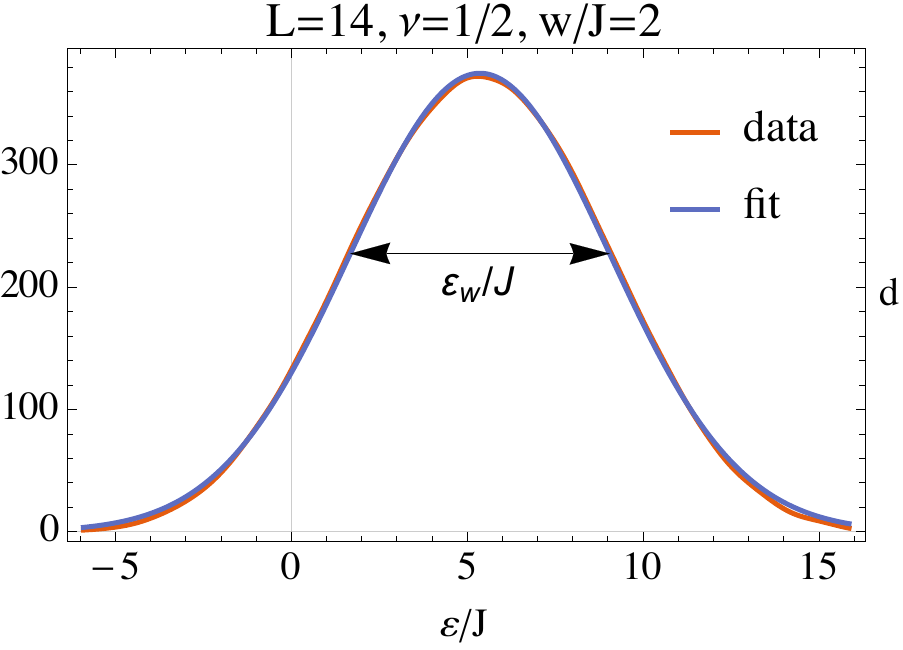}\protect\caption{Comparison between actual density of states and Gaussian fit.\label{fig:DOS-fit-example}}
\end{figure}

\begin{figure}
\includegraphics[scale=0.75]{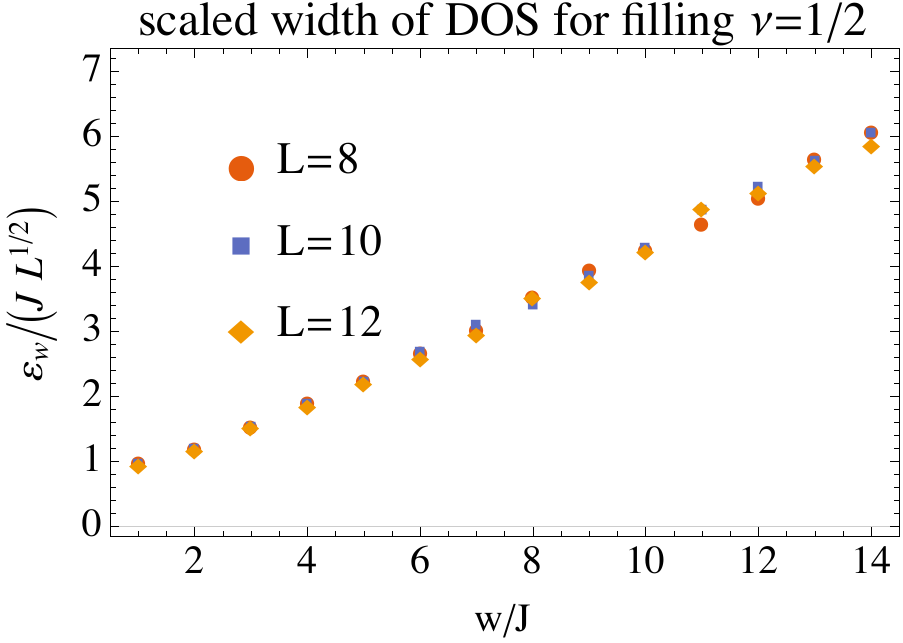}\protect\caption{Width of DOS as a function of disorder amplitude for two system sizes
$L=10,12$. This was obtained by explicitly calculating the DOS for
100,400 realisations of the disorder for $L=12,10$, respectively.\label{fig:Width-of-DOS}}
\end{figure}

The density of states (DOS) for a large class of systems with bounded local
Hilbert spaces, including the type we study, is Gaussian: $d\left(\epsilon\right)=\mathcal{N}^{-1}\exp\left(-\left(\epsilon-\epsilon_{0}\right)^{2}/2\epsilon_{w}^{2}\right)$~\cite{Canovi2011}.
An example for a particular disorder realisation of our HCB model
is shown in Fig.~\ref{fig:DOS-fit-example},
while a plot of the fitted width $\epsilon_{w}$ as a function of
disorder amplitude $w$ is shown in Fig.~\ref{fig:Width-of-DOS}.
Ref.~\cite{Canovi2011} shows that, in the absence of disorder, $\epsilon_{w}/J\propto L$,
while Fig.~\ref{fig:Width-of-DOS} suggests that in the presence
of strong disorder $\epsilon_{w}/J\propto L^{1/2}$. This may be understood
via the central limit theorem: for strong disorder, the system's eigenvalues
are approximately sums of uniformly distributed random numbers (the
random potential at each site), and the probability distribution of
a sum of $L$ uniformly distributed random numbers approaches a normal
distribution with width $L^{1/2}$. In any case, in Fig.~\ref{fig:example-level-stats}
we use the actual values of $\epsilon_{w}/J$ obtained by fitting
and averaging over a number of realisations.


\section{Driven Two-Level System\label{sec:Energy-absorption-by-TLS}}

\begin{figure}
\includegraphics[scale=0.6]{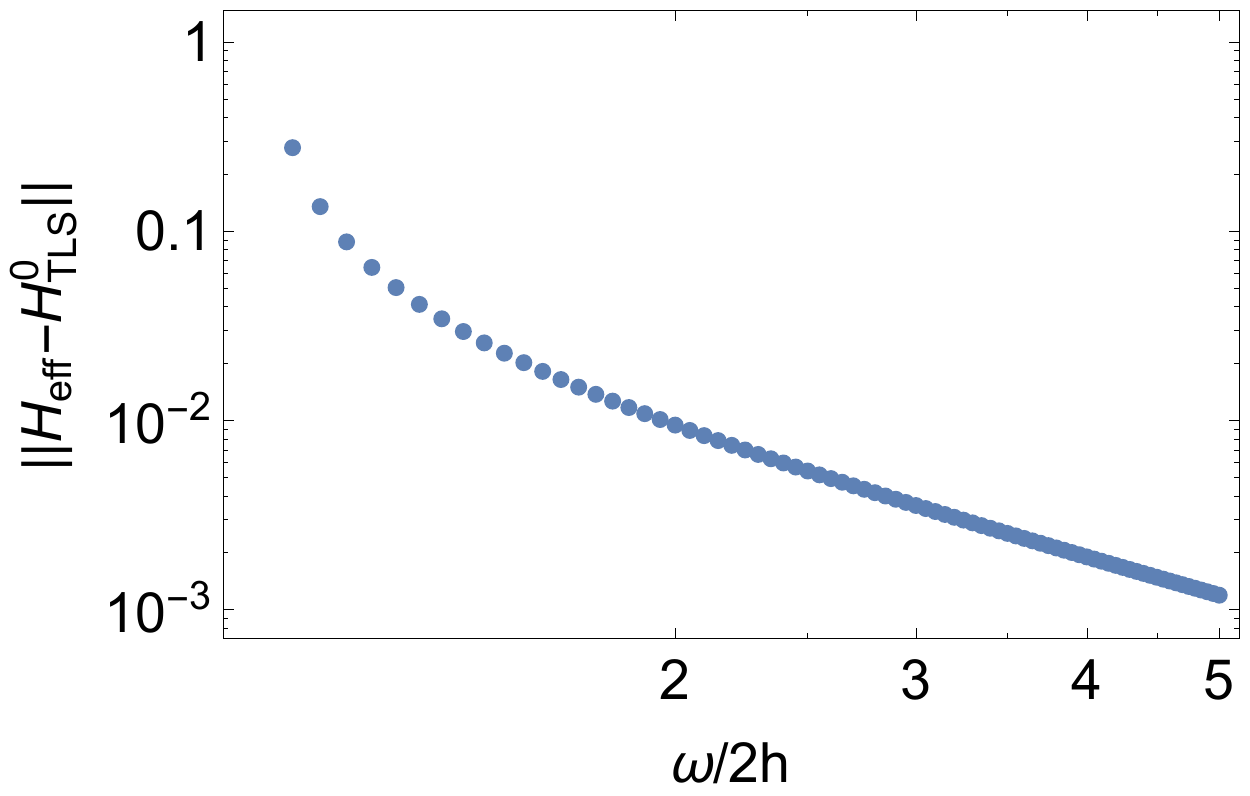}

\protect\caption{Approach of $H_{eff}$ to undriven $H_{TLS}$ with increasing frequency.
This is the limiting form for large $h/\delta$ (weak driving); in
practice, even $h/\delta=1$ gives a result indistinguishable from
this.\label{fig:frobenius-norm}}
\end{figure}

To illustrate how the energy bandwidth of a system sets a natural frequency
above which driving has little effect we consider a model with finite
bandwidth, namely the two-level system (TLS)~\cite{Grifoni1998}, driven with a small amplitude:
\begin{equation}
H_{TLS}=\begin{pmatrix}h & \delta\cos\left(\omega t\right)\\
\delta\cos\left(\omega t\right) & -h
\end{pmatrix}.\label{eq:htls}
\end{equation}
This has a threshold frequency above which it does not react to the
driving. The simplest way to see this is to directly calculate 
\[
\exp\left(-iH_{eff}\left(\epsilon\right)T\right)=\mathcal{T}\exp\left(-i\int_{\epsilon}^{\epsilon+T}H_{TLS}\left(\tau\right)d\tau\right)
\]
which describes the time evolution over one period. If $H_{eff}\approx\begin{pmatrix}h & 0\\
0 & -h
\end{pmatrix}=T^{-1}\int_{0}^{T}dt\, H_{TLS}\left(t\right)$ (compare to Eq.~\ref{eq:htls}) then clearly driving has very little
effect on the system. Fig.~\ref{fig:frobenius-norm} shows the Frobenius
norm $\left|H_{eff}-H_{TLS}^{0}\right|$ where $H_{TLS}^{0}$ is the
undriven TLS Hamiltonian (Eq.~\ref{eq:htls} with $\delta=0$).
This figure shows that the norm vanishes as a power law of $\omega/h$.

The situation for (for example) a single particle hopping on a lattice
(whether to nearest-neighbour sites or with some exponentially small
amplitude to hop to any distance) is completely identical.

The result described here is valid for weak driving amplitude; 
different physics might emerge in other limits~\cite{Das2010}.

\section{Delocalisation via dephasing\label{sec:Delocalisation-via-dephasing}}

\begin{figure}
\includegraphics[scale=0.4]{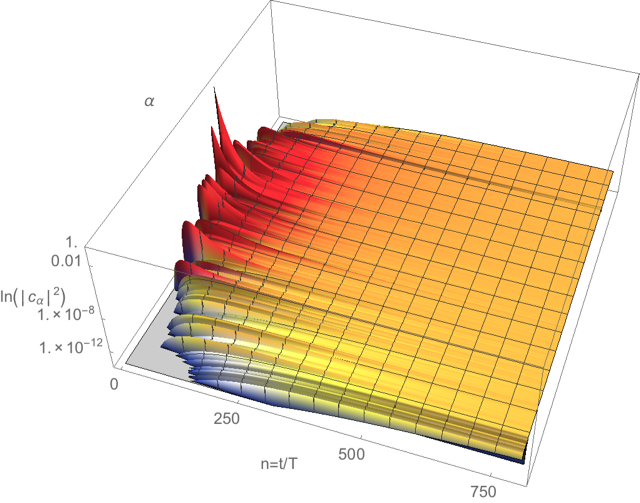}\protect\caption{Stroboscopic demonstration of the spreading of an initial energy-localised
state driven via alternating between two Hamiltonians and forcing
dephasing by hand (see Eqs.~3 and 4
in the main text for the Hamiltonians). The parameters used are system
size $L=10$ at half-filling, driving amplitude $\delta/J=0.2$, disorder
amplitude $w/J=8$ (inside the localised phase) and interactions $V_{1}/J=V_{2}/J=1$.
The system delocalises in energy even though both Hamiltonians are
localised as a direct result of dephasing.\label{fig:stosszahlansatz}}
\end{figure}

We show that an initial state which is an eigenstate of the Hamiltonian
$H_{1}$ and driven by alternating between $H_{1}$ and $H_{2}$ (as in the
protocol described in the main text for the local disordered model)
spreads out to fill the entire Hilbert space if it 
dephases between each change
of Hamiltonian. Specifically, we construct the density matrix $\rho\left(0\right)$,
diagonalise it in the eigenbasis of $H_{1}$ and then discard off-diagonal
elements. It is then rotated to the eigenbasis of $H_{2}$ and, again,
the off-diagonal terms are discarded. This procedure constitutes one
period. Discarding the off-diagonal states is supposed to model dephasing,
and is similar in spirit to the Boltzmann Stosszahlansatz. Figure~\ref{fig:stosszahlansatz}
shows a plot of $\ln\left(\left|c_{\alpha}\right|^{2}\right)$ (the
logarithimic scale is necesary as $\left|c_{\alpha}\right|^{2}$ ranges
from 1 to $1/D_{H}^{2}$ during the process, with $D_{H}$ the dimensions
of Hilbert space), with $c_{\alpha}$ the projection of the state
onto the eigenstates of $H_{1}$, as a function of period for an initial
state that as an eigenstate. The time evolution is carried out using
the driving protocol described in the main text for the local model
(Eqs.~3 and 4 in the article), with the additional
operation of dephasing carried out by hand (that is, off-diagonal
elements in the energy basis are discarded by hand). The paramerers
used are indicated in the caption.

The conclusion to be drawn from Fig.~\ref{fig:stosszahlansatz} is
that dephasing clearly leads to the system spreading out in energy
space. While not unexpected in general (a series of sudden perturbations
of the system will, in general, increase its energy), this calculation
confirms that the conclusion remains valid in the case of MBL Hamiltonians.

\section{The Quantum Random Energy Model\label{sec:The-QREM}}

Consider the set of $2^{N}$ Fock states for N spin-1/2 spins defined
by fixing all the $\sz[i]$ for $i=1\ldots N$, labelling them by
$\ket n$ with $n=1\ldots2^{N}$. These form a complete basis and
may be thought of as the ($2^{N}$) vertices of an N-dimensional hypercube.
To each edge/Fock state assign an energy at random, drawn from a distribution
\[
P\left(E\right)=\frac{1}{\sqrt{\pi N}}\exp\left(-E^{2}/N\right)
\]
(ensuring extensivity of the energies, bandwidth etc).

So far the problem is diagonal in the basis of the Fock states. Now
add a term $-\Gamma\sum_{i=1}^{N}\sx[i]$; the operator $\left(N-\sum_{i=1}^{N}\sx[i]\right)$
is the Laplacian on the hypercube so that $\Gamma$ plays the role
of a hopping amplitude. Overall, the Hamiltonian is the Anderson
problem with hopping $\Gamma$ and on an $N$-dimensional hypercube.
Formally, the Hamiltonian is given by
\[
\hat{H}=-\Gamma\hat{\nabla}^{2}+\hat{V}
\]
with $\hat{V}$ the random potential. This model is
not local in real space, but it is local on the hypercube, ie, the state
space of a spin-1/2 model.

\end{document}